\documentclass[pdflatex,sn-nature]{sn-jnl}

\usepackage{graphicx}%
\usepackage{multirow}%
\usepackage{amsmath,amssymb,amsfonts}%
\usepackage{amsthm}%
\usepackage{mathrsfs}%
\usepackage[title]{appendix}%
\usepackage{xcolor}%
\usepackage{textcomp}%
\usepackage{manyfoot}%
\usepackage{booktabs}%
\usepackage{algorithm}%
\usepackage{algorithmicx}%
\usepackage{algpseudocode}%
\usepackage{listings}%
\usepackage{adjustbox}%
\usepackage{makecell}  
\usepackage{float}

\theoremstyle{thmstyleone}%
\theoremstyle{thmstyletwo}%

\theoremstyle{thmstylethree}%

\begin{document}

\title[Article Title]{Multimodal Digital Biomarker for Asthma: Complementary Roles of Vocal, Clinical and Demographic Factors}


\author*[1]{\fnm{Vladimir} \sur{Despotovic}}\email{vladimir.despotovic@lih.lu}

\author[2]{\fnm{Milena} \sur{Despotovic}}\email{milena.despotovic@lih.lu}

\author[4]{\fnm{Abir} \sur{Elbeji}}\email{abir.elbeji@lih.lu}

\author[1,3]{\fnm{Petr V.} \sur{Nazarov}}\email{petr.nazarov@lih.lu}

\author[4]{\fnm{Guy} \sur{Fagherazzi}}\email{guy.fagherazzi@lih.lu}

\affil*[1]{\orgdiv{Bioinformatics \& AI, Department of Medical Informatics}, \orgname{Luxembourg Institute of Health}, \orgaddress{\street{1 A-B Rue Thomas Edison}, \city{Strassen}, \postcode{1445}, \country{Luxembourg}}}

\affil[2]{\orgdiv{Translational Medicine Operations Hub}, \orgname{Luxembourg Institute of Health}, \orgaddress{\street{1 A-B Rue Thomas Edison}, \city{Strassen}, \postcode{1445}, \country{Luxembourg}}}

\affil[3]{\orgdiv{Multi-Omics Data Science, Department of Cancer Research}, \orgname{Luxembourg Institute of Health}, \orgaddress{\street{1 A-B Rue Thomas Edison}, \city{Strassen}, \postcode{1445}, \country{Luxembourg}}}

\affil[4]{\orgdiv{Deep Digital Phenotyping, Department of Precision Health}, \orgname{Luxembourg Institute of Health}, \orgaddress{\street{1 A-B Rue Thomas Edison}, \city{Strassen}, \postcode{1445}, \country{Luxembourg}}}


\abstract{Asthma affects over 260 million people worldwide, yet diagnosis remains dependent on spirometry and specialist assessment, limiting accessibility in primary care and low-resource settings. Vocal biomarkers offer a promising non-invasive alternative, but prior studies have largely focused on acoustic features without integrating clinical context. We present a multimodal Mixture-of-Experts framework for asthma identification that adaptively combines acoustic embeddings from sustained vowel phonation and reading passage tasks with structured clinical and demographic data. The model was evaluated on a matched cohort of 1,218 self-reported asthma cases and healthy controls from the Colive Voice study. The multimodal model achieved an AUROC of 0.83 and Brier score of 0.18, outperforming unimodal approaches. Exploratory analysis of the gating mechanism in asthma cases showed that greater respiratory symptom burden was associated with increased weighting of reading-passage modality and reduced weighting of sustained-vowel phonation modality. These findings support the feasibility of voice-based identification of self-reported asthma; however, independent prospective validation remains necessary before clinical use.}

\keywords{asthma, voice, digital biomarker, mixture-of-experts}



\maketitle

\section{Introduction}\label{sec1}

Asthma is a chronic inflammatory respiratory condition affecting 262 million people worldwide and represents the second leading cause of mortality among chronic respiratory conditions \cite{Vos2020, Soriano2020}. Despite declining prevalence rates over recent decades, the absolute number of cases continues to rise, and the disease imposes substantial morbidity, disability-adjusted life years, and economic burden \cite{Wang2023, Yuan2025}. Early and accurate detection is essential for effective management and prevention of exacerbations; however, asthma diagnosis remains a global clinical challenge \cite{Venkatesan2025, Louis2022}. Standard assessment relies on spirometry to demonstrate reversible airflow limitation, yet this approach faces two compounding limitations: access to quality-assured spirometry is consistently constrained in primary care and low-resource settings \cite{Hakizimana2024}, and clinic-based spirometry may yield normal results due to fluctuating phases of symptoms, contributing to widespread under-diagnosis and delayed treatment \cite{Myers2025}. These barriers have motivated growing interest in scalable, non-invasive digital biomarkers that can support screening and monitoring outside traditional healthcare environments \cite{Chan2025, Kaur2023, Pizzimenti2026}.

The human voice has emerged as a promising source of digital biomarkers in multiple respiratory conditions, including asthma, chronic obstructive pulmonary disease, and COVID-19 \cite{Kaur2023, Xia2022, Yan2025, Han2022, Despotovic2021}. Asthma-related airway inflammation, bronchoconstriction, and mucus hypersecretion can alter subglottal airflow dynamics and expiratory control, with downstream effects on vocal fold vibration, phonatory stability, and connected speech production. A recent case-control study achieved an accuracy of 0.81 for asthma detection using acoustic features from sustained vowel phonation, underscoring the discriminative potential of voice-based approaches \cite{Lyu2025}. 

The complementary information of different voice tasks has been largely unexploited in respiratory disease research. Sustained vowel phonation provides a controlled window into phonatory quality, capturing jitter, shimmer, harmonics-to-noise ratio and maximum phonation time, without articulatory or prosodic confounds \cite{Fagherazzi2021}. Reading passage, by contrast, engages connected speech, capturing breath-group patterns, pause behaviour, speech rate, and articulatory dynamics that reflect respiratory-phonatory coordination under continuous subglottal demand \cite{Gerratt2016}. These tasks are therefore complementary rather than redundant: the former isolates laryngeal function and breath support, the latter probes respiratory patterning during natural speech. This complementarity has already been demonstrated in respiratory quality of life prediction, where fusing sustained vowel phonation and reading tasks using cross-attention improved accuracy by up to 4.2\% compared to either task alone \cite{Despotovic24_interspeech}. 

At the same time, demographic and clinical factors, including age, sex, body mass index (BMI), smoking history and comorbidities, play a well-established role in asthma risk, severity, and phenotypic heterogeneity \cite{Zein2021}. Machine learning models trained on structured clinical data alone have demonstrated meaningful, if moderate, discriminative performance for asthma detection and exacerbation prediction \cite{Molfino2024, Budiarto2023}. Integrating these structured data with voice-derived features offers an opportunity to improve predictive performance and clinical relevance through complementary information capture \cite{Despotovic2024}. Multimodal learning frameworks are well suited to this problem; however, effectively combining heterogeneous modalities remains technically challenging, particularly when their predictive contributions differ in scale and reliability. A further underappreciated challenge is the absence of large, well-labelled asthma-specific voice datasets, which limits the generalisability of models trained on voice data alone and strengthens the motivation for augmenting acoustic features with clinical annotations.

Mixture-of-Experts (MoE) architecture provides a principled approach to multimodal integration by learning to dynamically weight multiple specialised predictors. MoE has become a cornerstone of vision-language models, including MoE-LLaVA \cite{Lin2026}, MoME \cite{Shen2024}, and DeepSeek-VL2 \cite{wu2024}, where expert modules specialise over visual and linguistic feature subspaces and gating network learns to route inputs adaptively across modalities. The same principle applies in the clinical multimodal setting, where modalities such as acoustic voice embeddings and structured clinical variables are similarly heterogeneous in their representational space and predictive properties. However, unlike vision-language models, where MoE is deployed primarily as a computational scaling mechanism over large-scale data, clinical applications demand that gating network remains robust and interpretable under small, imbalanced, and noisy datasets, and that expert utilisation patterns yield clinically meaningful insights into modality contributions at the participant level. This enables the model to account for variability in data quality and individual participant characteristics, and can improve both predictive performance and interpretability. MoE architectures have been recently explored in general clinical prediction tasks using multiple modalities, including electronic health records, clinical notes and imaging data \cite{Wang25, Gao2026}. Nevertheless, their application to clinical  biomarker problems, and to multimodal vocal biomarkers in particular, remains largely unexplored.

In this study, we propose a multimodal MoE framework for self-reported asthma prediction that integrates acoustic embeddings derived from two complementary voice tasks — sustained vowel phonation and reading passage — with clinical and demographic features. The model learns to combine modality-specific experts through a gating network, enabling adaptive fusion of information at the participant level. Our contributions are threefold: (1) we demonstrate the feasibility of combining multiple voice tasks with clinical and demographic data for self-reported asthma prediction in a unified learning framework; (2) we introduce a MoE architecture tailored to multimodal digital biomarkers, explicitly leveraging the distinct acoustic properties of sustained phonation and connected speech; and (3) we provide insights into modality-specific contributions through analysis of the gating mechanism. This work highlights the potential of multimodal AI to support accessible, non-invasive respiratory disease assessment. 

\begin{table}[t]
\centering
\caption{Cohort characteristics}
\label{tab:cohort}
\footnotesize
\setlength{\tabcolsep}{4pt}
\begin{tabular}{lllcc>{\centering\arraybackslash}p{1.5cm}}
\toprule
\multicolumn{3}{l}{} & \textbf{Healthy controls} & \textbf{Asthma} & \textbf{p value} \\
\midrule
\multicolumn{3}{l}{\textbf{Number of recordings}} & 609 & 609 & \\
\midrule

\multirow{10}{*}[-7.2pt]{\textbf{Demographic}}
    & \multirow{2}{*}{\textbf{Sex}}
        & Female & 419 (68.8) & 419 (68.8) & \multirow{2}{*}{1} \\
    &   & Male   & 190 (31.2) & 190 (31.2) & \\
\cmidrule(l){2-6}
    & \multirow{2}{*}{\textbf{Age}}
        & Female & $43.1 \pm 14.6$ & $43.1 \pm 14.6$ & \multirow{2}{*}{0.94} \\
    &   & Male   & $39.8 \pm 13.7$ & $40.0 \pm 14.3$ & \\
\cmidrule(l){2-6}
    & \multirow{2}{*}{\textbf{Language}}
        & English & 371 (60.9) & 371 (60.9) & \multirow{2}{*}{1} \\
    &   & French  & 238 (39.1) & 238 (39.1) & \\
\cmidrule(l){2-6}
    & \multirow{4}{*}{\textbf{BMI$^{1}$}}
        & Underweight & 24 (3.9)   & 21 (3.4)   & \multirow{4}{*}{\textbf{$<$0.001}} \\
    &   & Normal      & 308 (50.6) & 233 (38.3) & \\
    &   & Overweight  & 161 (26.4) & 158 (25.9) & \\
    &   & Obesity     & 116 (19.1) & 197 (32.3) & \\
\midrule

\multirow{18}{*}{\textbf{Lifestyle}}
    & \multirow{3}{*}{\textbf{Smoking}}
        & Not at all      & 529 (86.9) & 498 (81.8) & \multirow{3}{*}{\textbf{0.006}} \\
    &   & Less than daily & 33 (5.4)   & 30 (4.9)   & \\
    &   & Daily           & 47 (7.7)   & 81 (13.3)  & \\
\cmidrule(l){2-6}
    & \multirow{10}{*}{\textbf{Alcohol}}
        & Never                & 80 (13.1)  & 130 (21.3) & \multirow{10}{*}{\textbf{$<$0.001}} \\
    &   & 1--2 times per year  & 47 (7.7)   & 65 (10.7)  & \\
    &   & 3--11 times per year & 68 (11.2)  & 85 (14.0)  & \\
    &   & 1 time per month     & 43 (7.1)   & 37 (6.1)   & \\
    &   & 2--3 times per month & 83 (13.6)  & 77 (12.6)  & \\
    &   & 1 time per week      & 78 (12.8)  & 54 (8.9)   & \\
    &   & 2 times per week     & 100 (16.4) & 78 (12.8)  & \\
    &   & 3--4 times per week  & 66 (10.8)  & 45 (7.4)   & \\
    &   & 5--6 times per week  & 30 (4.9)   & 24 (3.9)   & \\
    &   & Everyday             & 14 (2.3)   & 14 (2.3)   & \\
\cmidrule(l){2-6}
    & \multirow{5}{*}{\textbf{Stress}}
        & Not at all     & 126 (20.7) & 90 (14.8)  & \multirow{5}{*}{\textbf{$<$0.001}} \\
    &   & A little       & 257 (42.2) & 200 (32.8) & \\
    &   & To some extent & 160 (26.3) & 154 (25.3) & \\
    &   & Rather much    & 51 (8.4)   & 110 (18.1) & \\
    &   & Very much      & 15 (2.5)   & 55 (9.0)   & \\
\midrule

\multirow{2}{*}{\textbf{Treatment}}
    & \multirow{2}{*}{\textbf{Antihistamines}}
        & No  & 584 (95.9) & 470 (77.2) & \multirow{2}{*}{\textbf{$<$0.001}} \\
    &   & Yes & 25 (4.1)   & 139 (22.8) & \\
\midrule

\multirow{2}{*}{\textbf{Symptoms}}
    & \multicolumn{2}{l}{\textbf{RQoL}$^{2}$} & $14.5 \pm 2.9$  & $21.4 \pm 8.3$  & \textbf{$<$0.001} \\
    & \multicolumn{2}{l}{\textbf{FSS}$^{3}$}  & $28.8 \pm 10.8$ & $37.8 \pm 13.5$ & \textbf{$<$0.001} \\
\bottomrule
\multicolumn{6}{l}{} \\
\multicolumn{6}{l}{%
  \begin{minipage}{1.0\textwidth}
  \footnotesize
  Values for categorical variables are reported as n (\%);
  continuous variables as mean $\pm$ standard deviation.
  Bold p values indicate statistical significance (p $<$ 0.05). \\[2pt]
  $^{1}$BMI: Body Mass Index (kg/m$^{2}$); categories defined according to WHO
  classification: Underweight $<$18.5, Normal 18.5--24.9, Overweight 25.0--29.9,
  Obesity $\geq$30.0. \\[2pt]
  $^{2}$RQoL: Respiratory Quality of Life, assessed using the VQ11
  questionnaire (range 11--55); higher scores indicate worse RQoL. \\[2pt]
  $^{3}$FSS: Fatigue Severity Scale (range 9--63); higher scores indicate
  greater fatigue severity.
  \end{minipage}%
} \\
\end{tabular}
\end{table}

\section{Results}\label{sec2}

\subsection{Dataset and cohort characteristics}\label{dataset}

The study cohort comprised 1,218 participants drawn from the Colive Voice\footnote{https://www.colivevoice.org} study, equally distributed between individuals with self-reported asthma diagnosis and healthy controls, matched by sex, age and language (Table~\ref{tab:cohort}). Healthy controls were required to have no history of chronic respiratory or neurological conditions known to affect voice production, ensuring that observed acoustic differences between groups could be attributed to asthma rather than confounding pathologies. Recordings were collected in English (n=742) and French (n=476).

The two groups differed significantly on several clinical and lifestyle variables. The asthma group showed a notably higher prevalence of obesity and a lower proportion of participants with normal body mass index (BMI), resulting in a significant difference in BMI distribution (p$<$0.001). Daily smoking was more prevalent among participants with asthma (p=0.006), and antihistamine use was substantially higher in the asthma group (p$<$0.001), consistent with the frequent co-occurrence of asthma and allergic conditions \cite{Hossny2024}. Participants with asthma also reported greater perceived stress (p$<$0.001) and differed in self-reported alcohol consumption patterns (p$<$0.001).
Reflecting the broader health impact of the disease, participants with asthma reported significantly worse Respiratory Quality of Life (RQoL; p$<$0.001) and Fatigue Severity Score (FSS; p$<$0.001) compared to healthy controls, underscoring the systemic burden of asthma beyond airway function alone.

\begin{figure}[h]
\centering
\includegraphics[width=1.0\textwidth]{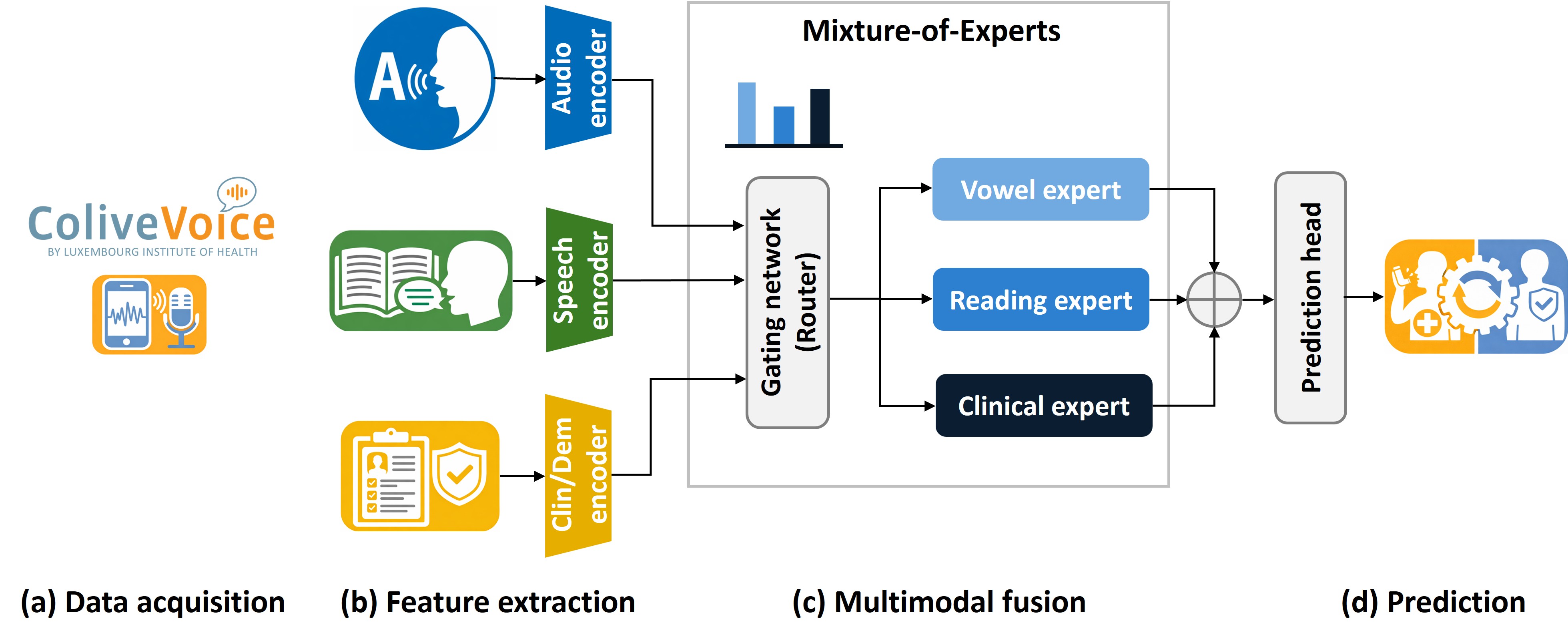}
\caption{Multimodal Mixture-of-Experts framework for self-reported asthma identification from sustained vowel phonation, reading passage and clinical/demographic data; 
(a) Data acquisition: participants provide sustained vowel phonation and reading passage recordings via an online application, alongside clinical and demographic data collected using clinically validated questionnaires;
(b) Feature extraction: task-specific voice encoders process vowel and continuous speech signals, while a separate encoder processes clinical/demographic inputs;
(c) Multimodal fusion: a gating network (router) adaptively weights modality-specific representations and routes them to expert subnetworks (vowel, reading, and clinical experts). For clarity, a single expert per modality is illustrated; in practice, each modality may be associated with multiple experts;
(d) Prediction: expert outputs are aggregated and passed to a prediction head to generate the final classification between self-reported asthma cases and healthy controls.}
\label{fig:MoE}
\end{figure}

\begin{figure}[b]
\centering
\includegraphics[width=1.0\textwidth]{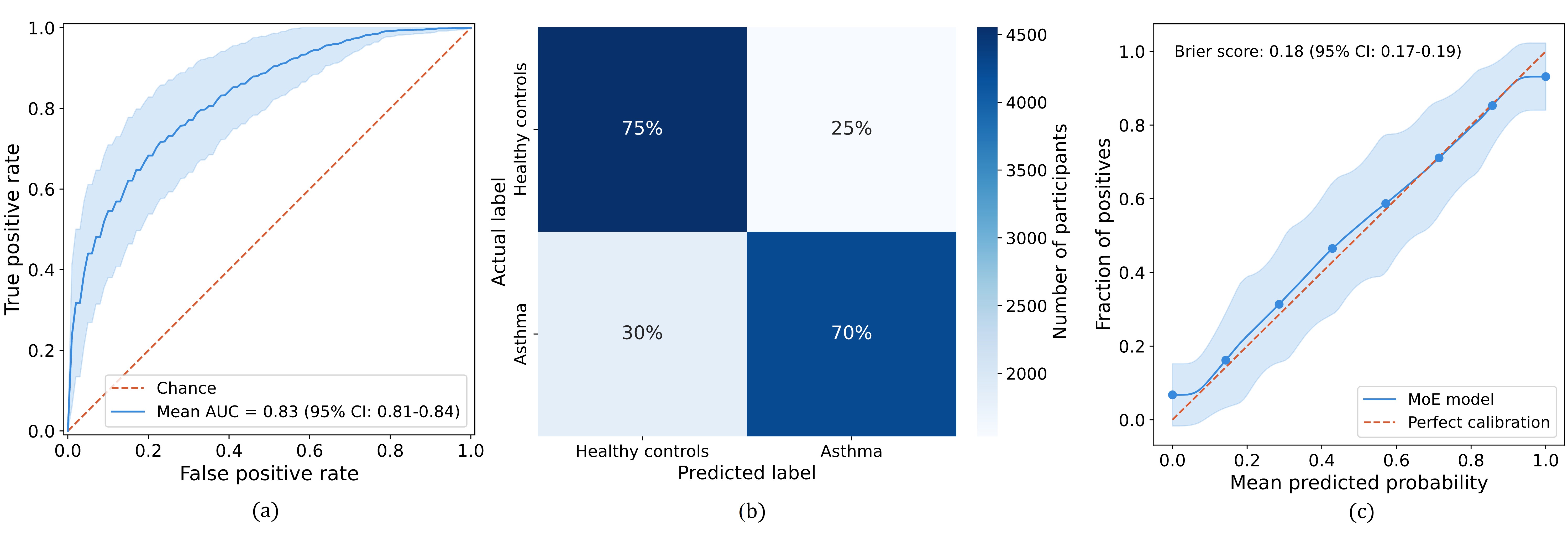}
\caption{Predictive performance of the full multimodal MoE model for self-reported asthma identification.
(a) ROC curve averaged across repeated 10$\times$10 stratified cross-validation folds (solid blue line) with standard deviation band (shading), individual fold curves (light blue lines) and AUROC reported as mean with 95\% confidence intervals; (b) Aggregated confusion matrix across all folds; (c) Calibration curve averaged across all folds (solid blue line)  with standard deviation band (shading) and Brier score reported as mean with 95\% confidence intervals. \\MoE: Mixture-of-Experts; ROC: Receiver Operating Characteristic; AUROC: Area under the ROC curve; HC: healthy controls. }
\label{fig:performance}
\end{figure}

\subsection{Multimodal self-reported asthma identification via Mixture-of-Experts fusion}\label{MoE}

We developed a multimodal MoE framework for self-reported asthma identification that integrates acoustic embeddings derived from two standardised voice tasks — sustained vowel phonation and reading passage — with structured clinical and demographic features, as shown in Figure~\ref{fig:MoE}. Each modality is processed by a set of dedicated expert networks: four sustained vowel experts (VE1--VE4), four reading passage experts (RE1--RE4), and two clinical/demographic experts (CE1--CE2), yielding ten experts that learn complementary representations of the input. A gating network, conditioned jointly on all three modalities, learns to assign adaptive weights to each expert for every individual input, enabling the model to dynamically adjust its reliance on voice-derived and clinical/demographic information depending on the participant characteristics. The final prediction is computed as a weighted mixture of expert outputs, with the gating weights providing a transparent record of modality contributions. The model was trained and evaluated using repeated 10$\times$10 stratified cross-validation on a matched cohort of self-reported asthma cases and healthy controls, following the recommendations in \cite{deHond2023}. Performance was assessed across discrimination, calibration, and modality contribution metrics.

The full multimodal MoE model, integrating sustained vowel phonation, reading passage, and clinical/demographic features, achieved a mean Area Under the Receiver Operating Characteristic curve (AUROC) of 0.83 (95\% CI: 0.81--0.84) (Figure~\ref{fig:performance}(a)), indicating robust discriminative performance with low variability across folds. The model correctly identified 70\% of self-reported asthma cases (sensitivity) and 75\% of healthy controls (specificity), demonstrating nearly balanced classification performance (Figure~\ref{fig:performance}(b)). Overall accuracy, precision, and F1 score were 0.72, 0.75 and 0.71, respectively.

Model calibration was assessed using the Brier score and a calibration curve in Figure~\ref{fig:performance}(c). The mean Brier score of 0.18 (95\% CI: 0.17--0.19) indicates adequate probabilistic calibration. The calibration curve closely follows the diagonal representing perfect calibration, suggesting that predicted probabilities are generally well aligned with observed outcome frequencies. Slight underestimation of risk was observed across low-to-mid probability range. 

\subsection{Modality-level routing analysis}\label{interpretability}

The gating mechanism of the MoE model assigns adaptive weights to each expert based on the combined representation of audio and clinical inputs, allowing the routing of information across modalities to be examined at both the expert and modality-group level.

\begin{figure}[h]
\centering
\includegraphics[width=1.0\textwidth]{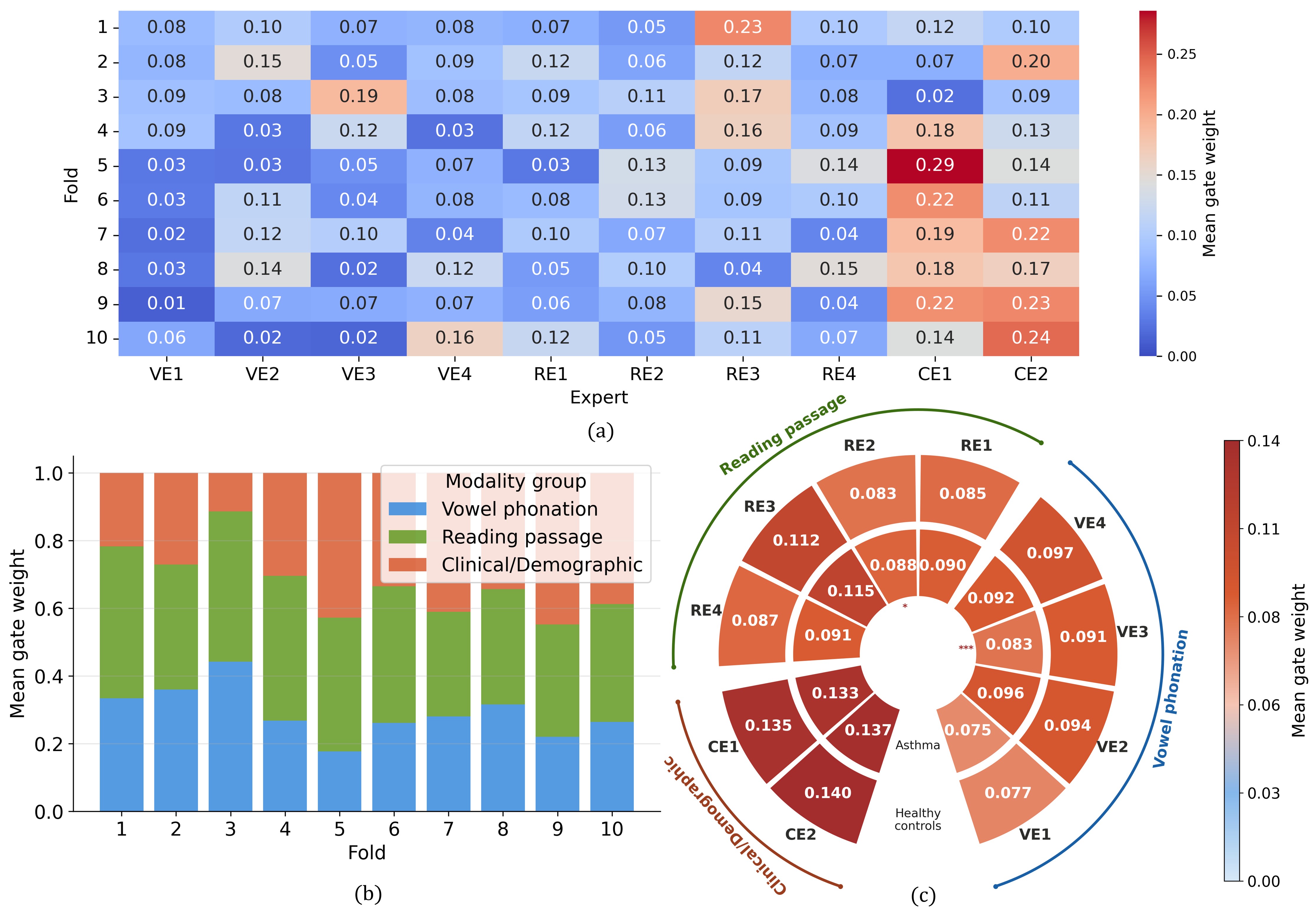}
\caption{Modality-level routing analysis of the full multimodal MoE model.
(a) Expert gate weights across ten stratified cross-validation folds. Experts are labelled by modality group: VE1–-VE4 (sustained vowel phonation), RE1–-RE4 (reading passage), CE1–-CE2 (clinical/demographic data). Colour encodes mean gate weight from low (blue) to high (red); (b) Modality-group gate weights per fold, shown as stacked bars normalised to 1.0; (c) Expert gate weights averaged across all participants, stratified by class. Inner ring: self-reported asthma cases; outer ring: healthy controls. Colour encodes mean gate weight from low (blue) to high (red). Asterisks indicate significant differences between groups (Mann-Whitney U test): * $p<0.05$, ** $p<0.01$, *** $p<0.001$.}
\label{fig:interpretability}
\end{figure}

Analysis of expert-level gate weights across ten cross-validation folds revealed data-dependent routing behaviour, with no single expert dominating consistently across all folds, as shown in Figure~\ref{fig:interpretability}(a). Active experts, defined as those receiving non-negligible gate weight, spanned all three modality groups, confirming that no modality was systematically suppressed. Despite variability, several consistent patterns emerged. Both clinical/demographic experts received consistently high gate weights across folds, with CE1 peaking at 0.29 in fold 5 and CE2 showing sustained high weights in folds 7–10, suggesting that clinical and demographic features captured stable and informative patterns. In contrast, certain experts, (e.g. VE1, RE1), were frequently assigned lower weights across folds, indicating potential redundancy or limited utility in the learned representation.

Modality-group contributions, shown in Figure~\ref{fig:interpretability}(b), show no strong dominance of any single modality, with notable variability across folds. Reading passage remained relatively stable, whereas clinical/demographic features and vowel phonation showed substantial fold-to-fold variability, suggesting that the gating mechanism adaptively redistributes weights depending on the data encountered in each fold.

At the participant level, expert gate weights were first averaged across all repetitions and cross-validation folds for each participant individually, yielding a stable per-participant gate weight vector. These vectors were then stratified by diagnostic class and averaged across participants within each class for visualisation in the circular heatmap in Figure~\ref{fig:interpretability}(c). Clinical/demographic experts CE1 and CE2 received the highest mean gate weights overall in both diagnostic classes, indicating that clinical and demographic features remain highly informative regardless of diagnosis. After correction across all ten expert-level comparisons (Benjamini-Hochberg), statistically significant differences between self-reported asthma cases and healthy controls were observed for sustained vowel phonation expert VE3 ($p<0.001$) and reading passage expert RE2 ($p=0.011$), suggesting that the gating mechanism differentially routes voice information depending on diagnostic class. No significant differences were observed among any other experts, including both clinical/demographic experts.

\begin{figure}[t]
\centering
\includegraphics[width=1.0\textwidth]{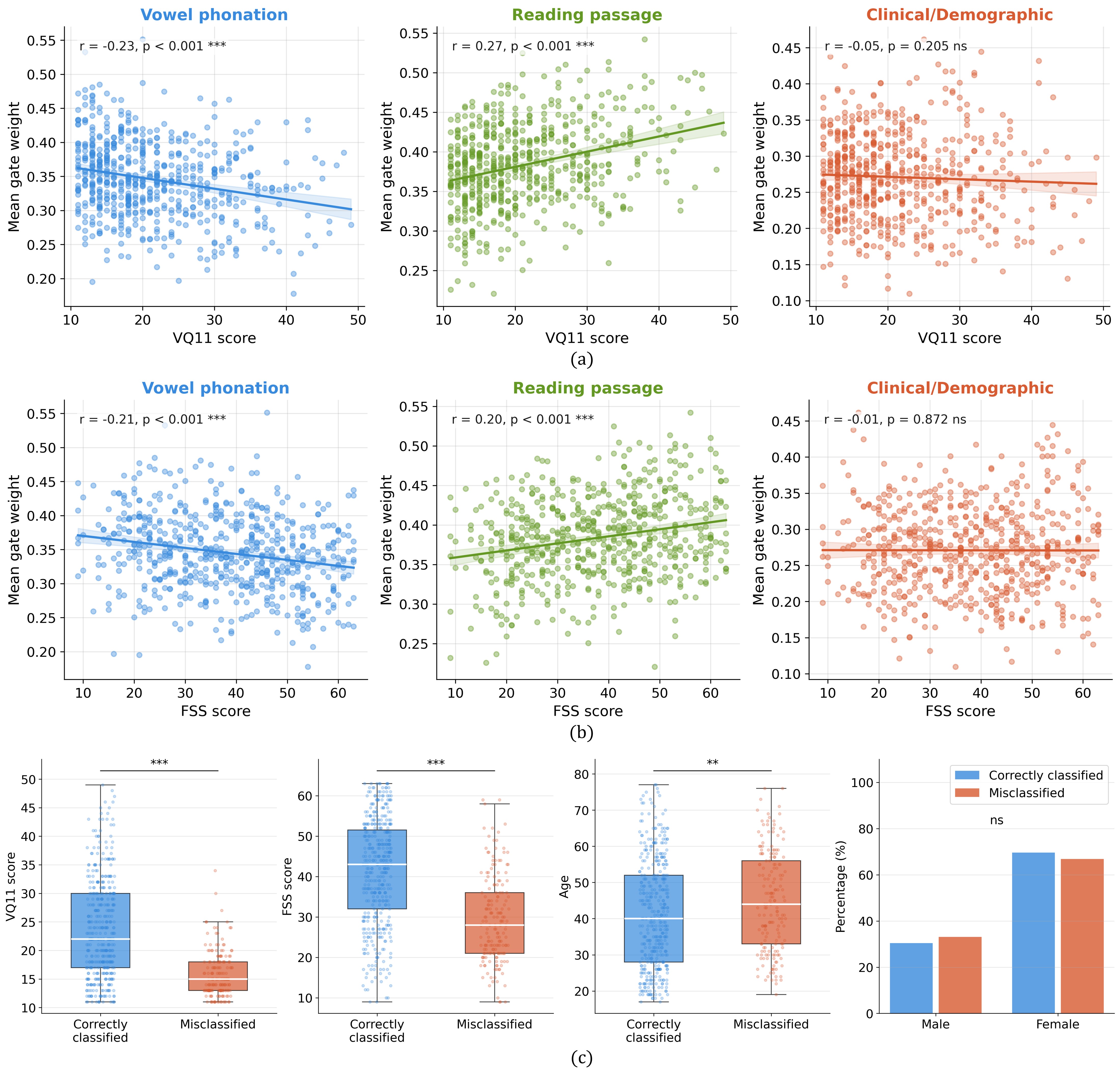}
\caption{Clinical correlates of gating behaviour and misclassification analysis in self-reported asthma cases. (a) Spearman correlation between modality group gate weights and respiratory quality of life score (VQ11) regression lines with 95\% confidence bands; (b) Spearman correlation between modality group gate weights and fatigue severity score (FSS) regression lines with 95\% confidence bands; (c) Clinical profiles of correctly classified (true positive, TP) and misclassified (false negative, FN) self-reported asthma cases, compared across VQ11 score, FSS score, age, and sex. Statistical significance for continuous variables (VQ11, FSS, and age) was assessed using the Mann–Whitney U test, while differences in sex distribution were assessed using the chi-square test of independence: * $p<0.05$, ** $p<0.01$, *** $p<0.001$, ns = not significant.}
\label{fig:clin_correlates}
\end{figure}

To further characterise the clinical relevance of the gating mechanism, modality group gate weights were correlated with continuous measures of symptom severity in self-reported asthma cases. 
Vowel phonation gate weights showed a significant negative correlation with respiratory quality of life (VQ11) score ($r=-0.23, p<0.001$), whereas reading passage gate weights showed a significant positive correlation ($r=0.27, p<0.001$), as shown in Figure~\ref{fig:clin_correlates}(a). A similar pattern was observed for fatigue severity score (FSS) (Figure~\ref{fig:clin_correlates}b): while vowel phonation gate weights were significantly negatively correlated with FSS ($r=-0.21, p<0.001$), reading passage gate weights were significantly positively correlated ($r=0.20, p<0.001$). Clinical/demographic gate weights showed no significant association with either VQ11 or FSS. 

To characterise model errors, clinical profiles of correctly classified (true positives) and misclassified asthma cases (false negatives) were compared in Figure~\ref{fig:clin_correlates}(c). Misclassified cases had significantly lower VQ11 scores ($p<0.001$) and FSS scores ($p<0.001$), and were significantly older ($p=0.004$) compared to correctly classified participants, indicating that the model was less sensitive to asthma cases characterised by lower symptom burden and older age. No significant difference was observed in the proportion of male and female participants between the two groups.

\subsection{Modality and feature contributions}\label{ablation}

An ablation study was conducted to assess the relative contribution of each modality and to quantify the performance gains achieved through multimodal integration (Table~\ref{tab:ablation}). Among unimodal models, clinical and demographic (CD) features yielded the strongest performance (AUROC 0.75, accuracy 0.69), outperforming both sustained vowel (SV) and reading passage (RP) inputs. RP alone showed the lowest overall performance (AUROC 0.64, accuracy 0.60), indicating limited discriminative power when used in isolation.

\begin{table}[t]
\caption{Model performance across unimodal, multimodal and baseline configurations.}
\label{tab:ablation}
\tiny
\setlength{\tabcolsep}{2pt}
\centering
\begin{tabular}{llllllll}
\toprule
\textbf{Model} & \multicolumn{1}{c}{\textbf{AUROC}} & \multicolumn{1}{c}{\textbf{Accuracy}} & \multicolumn{1}{c}{\textbf{Sensitivity}} & \multicolumn{1}{c}{\textbf{Specificity}} & \multicolumn{1}{c}{\textbf{Precision}} & \multicolumn{1}{c}{\textbf{F1 score}} & \multicolumn{1}{c}{\textbf{Brier score}} \\
\midrule
\multicolumn{8}{l}{\textit{Unimodal}} \\
SV & 0.69 (0.69--0.69)$^{*}$ & 0.63 (0.62--0.63)$^{*}$ & 0.65 (0.64--0.65)$^{*}$ & 0.61 (0.60--0.61)$^{*}$ & 0.62 (0.62--0.63)$^{*}$ & 0.63 (0.63--0.64)$^{*}$ & 0.23 (0.23--0.23)$^{*}$ \\
RP & 0.64 (0.64--0.64)$^{*}$ & 0.60 (0.60--0.60)$^{*}$ & 0.57 (0.56--0.57)$^{*}$ & 0.63 (0.63--0.64)$^{*}$ & 0.61 (0.60--0.61)$^{*}$ & 0.59 (0.58--0.59)$^{*}$ & 0.24 (0.24--0.24)$^{*}$ \\
CD & 0.75 (0.75--0.75)$^{*}$ & 0.69 (0.69--0.70)$^{*}$ & 0.65 (0.63--0.66)$^{*}$ & 0.74 (0.71--0.76) & 0.72 (0.71--0.73)$^{*}$ & 0.67 (0.67--0.68)$^{*}$ & 0.20 (0.20--0.20)$^{*}$ \\
\midrule
\multicolumn{8}{l}{\textit{Multimodal MoE}} \\
SV+RP & 0.71 (0.69--0.73)$^{*}$ & 0.63 (0.62--0.65)$^{*}$ & 0.66 (0.63--0.69)$^{*}$ & 0.61 (0.59--0.62)$^{*}$ & 0.64 (0.63--0.65)$^{*}$ & 0.64 (0.61--0.66)$^{*}$ & 0.23 (0.23--0.24)$^{*}$ \\
SV+CD & 0.81 (0.79--0.82) & 0.71 (0.69--0.72) & \textbf{0.70 (0.65--0.74)} & 0.72 (0.68--0.75) & 0.73 (0.71--0.74) & 0.70 (0.68--0.72) & 0.19 (0.18--0.20) \\
RP+CD & 0.81 (0.80--0.83) & 0.71 (0.70--0.72) & 0.64 (0.61--0.67)$^{*}$ & \textbf{0.79 (0.75--0.83)} & \textbf{0.77 (0.75--0.80)} & 0.68 (0.67--0.70) & 0.19 (0.18--0.19) \\
\textbf{SV+RP+CD} & \textbf{0.83 (0.81--0.84)} & \textbf{0.72 (0.70--0.74)} & \textbf{0.70 (0.67--0.73)} & 0.75 (0.72--0.78) & 0.75 (0.73--0.78) & \textbf{0.71 (0.69--0.73)} & 0.18 (0.17--0.19) \\
\midrule
\multicolumn{8}{l}{\textit{Baselines}} \\
LR & 0.77 (0.77--0.78)$^{*}$ & 0.69 (0.69--0.69)$^{*}$ & 0.69 (0.68--0.70) & 0.69 (0.69--0.69)$^{*}$ & 0.69 (0.69--0.70)$^{*}$ & 0.69 (0.69--0.70) & 0.20 (0.20--0.20)$^{*}$ \\
MLP & 0.74 (0.74--0.74)$^{*}$ & 0.67 (0.67--0.68)$^{*}$ & 0.68 (0.67--0.69) & 0.66 (0.66--0.66)$^{*}$ & 0.67 (0.66--0.67)$^{*}$ & 0.67 (0.67--0.68)$^{*}$ & 0.24 (0.24--0.25)$^{*}$ \\
XGBoost & 0.81 (0.81--0.82) & \textbf{0.72 (0.71--0.73)} & 0.65 (0.64--0.66)$^{*}$ & \textbf{0.79 (0.78--0.80)} & 0.76 (0.75--0.77) & 0.70 (0.69--0.71) & \textbf{0.17 (0.17--0.18)} \\
\bottomrule
\end{tabular}
\vspace{3pt}
\parbox{0.95\linewidth}{\footnotesize Values are means (95\% confidence intervals) across repeated 10$\times$10 stratified cross-validation. Baseline models were fitted to concatenated clinical/demographic, sustained vowel phonation and reading task features. Bold values indicate the best performance per metric. Where two configurations share the best performance, both are shown in bold. Asterisks indicate a significant difference from the full MoE model at $p<0.05$ (two-sided Wilcoxon signed-rank test with Benjamini--Hochberg false discovery rate correction across multiple comparisons within each metric). \\
SV: sustained vowel phonation; RP: reading passage; CD: clinical and demographic features; MoE: mixture-of-experts; LR: logistic regression; MLP: multilayer perceptron; XGBoost: extreme gradient boosting.}
\end{table}

Combining voice and clinical modalities consistently improved performance over unimodal configurations. The MoE models integrating SV+CD and RP+CD achieved identical and the highest performance among bimodal setups (AUROC 0.81), though with distinct trade-offs: SV+CD achieved a better balance between sensitivity and specificity (0.70 and 0.72, respectively), while RP+CD achieved higher specificity and precision (0.79 and 0.77, respectively) at the cost of lower sensitivity (0.64). In contrast, SV+RP yielded only modest gains over the voice unimodal models (AUROC 0.71, accuracy 0.63), performing similarly to SV alone, suggesting that clinical information contributes substantially more to bimodal performance than combining the two audio modalities.

The full multimodal MoE model (SV+RP+CD) achieved the best overall performance, with an AUROC of 0.83 (95\% CI: 0.81--0.84), accuracy of 0.72 (95\% CI: 0.70--0.74), and F1 score of 0.71 (95\% CI: 0.69--0.73). 

To assess whether the MoE architecture provides an advantage beyond simple concatenation of three feature sets, it was benchmarked against three conventional machine learning models. Logistic regression and Multilayer Perceptron (MLP) performed significantly worse than the full MoE across most of the metrics. XGBoost was the strongest baseline and the only model approaching MoE performance, with comparable AUROC (0.81) and accuracy (0.72). However, the MoE achieved significantly higher sensitivity (0.70 vs. 0.65, p = 0.012), and a more favorable sensitivity–specificity balance than XGBoost.

To assess the statistical significance of the differences in Table~\ref{tab:ablation}, each configuration was compared with the full multimodal model within each cross-validation repetition. Unimodal and voice-only bimodal configurations differed significantly from the full model in most of the metrics. In contrast, neither bimodal voice-plus-clinical configuration differed significantly from the full model on any metric.

Overall, these results highlight the complementary nature of audio and clinical data, with clinical features providing strong baseline predictive power and audio modalities contributing additional discriminative and calibration benefits when combined within the MoE framework. The modest incremental gain of the full model over the best bimodal configurations suggests a degree of redundancy between the two audio modalities when clinical features are already included.

\begin{table}[t]
\caption{Model performance by language, sex, and age group.}
\label{tab:subgroups}
\setlength{\tabcolsep}{6pt}
\centering
\begin{tabular}{lccccc}
\toprule
\textbf{Subgroup} & \multicolumn{1}{c}{\textbf{Participants}}
& \multicolumn{1}{c}{\textbf{AUROC}}
& \multicolumn{1}{c}{\textbf{Sensitivity}}
& \multicolumn{1}{c}{\textbf{Specificity}}
& \multicolumn{1}{c}{\textbf{Brier score}} \\
\midrule
\multicolumn{6}{l}{\textit{Language}} \\
\quad French & 476 & 0.84 & 0.71 & 0.77 & 0.17 \\
\quad English & 742 & 0.82 & 0.70 & 0.74 & 0.19 \\
\midrule
\multicolumn{6}{l}{\textit{Sex}} \\
\quad Female & 838 & 0.84 & 0.71 & 0.76 & 0.17 \\
\quad Male & 380 & 0.80 & 0.68 & 0.72 & 0.20 \\
\midrule
\multicolumn{6}{l}{\textit{Age}} \\
\quad $<$40 years & 609 & 0.82 & 0.74 & 0.72 & 0.18 \\
\quad $\geq$40 years & 609 & 0.84 & 0.66 & 0.79 & 0.18 \\
\bottomrule
\end{tabular}
\vspace{3pt}
\parbox{1.0\linewidth}{\footnotesize Values are derived from the same cross-validated predictions used to evaluate the full multimodal MoE model, averaged  across repeated 10$\times$10 stratified cross-validation folds.}
\end{table}

\begin{table}[b]
\caption{Cross-lingual transfer.}
\label{tab:cross-language}
\setlength{\tabcolsep}{20pt}
\centering
\begin{tabular}{lcc}
\toprule
\textbf{Metric}
& \multicolumn{1}{c}{\textbf{French $\rightarrow$ English}}
& \multicolumn{1}{c}{\textbf{English $\rightarrow$ French}} \\
\midrule
\textbf{AUROC} & 0.79 (0.76--0.82) & 0.80 (0.75--0.86) \\
\textbf{Sensitivity} & 0.84 (0.77--0.92) & 0.88 (0.81--0.94) \\
\textbf{Specificity} & 0.51 (0.39--0.64) & 0.51 (0.35--0.67) \\
\textbf{Brier score} & 0.21 (0.18--0.24) & 0.21 (0.16--0.25) \\
\bottomrule
\end{tabular}
\vspace{3pt}
\parbox{1.0\linewidth}{\footnotesize Models were trained on one language and evaluated on the other, with language excluded from the clinical feature set. Each direction was retrained ten times with different random seeds. Reported values are means with 95\% confidence intervals.}
\end{table}

Model performance was further examined separately by language, sex, and age group, using the same cross-validated predictions as in Table~\ref{tab:ablation}. Model performance was broadly consistent across subgroups (Table~\ref{tab:subgroups}), with AUROC ranging from 0.80 to 0.84, and slightly higher discrimination observed in French-speaking participants, females, and participants aged over 40 years. Sensitivity and specificity showed modest variation across language and sex subgroups. In age subgroups, however, sensitivity was notably lower in participants aged 
above 40 years (0.66) compared to those under 40 (0.74), while specificity showed the opposite pattern (0.79 versus 0.72), indicating that the model was less able to correctly identify self-reported asthma cases among older participants. This is consistent with the error analysis, which showed that misclassified asthma cases were significantly older than correctly classified cases.

Calibration varied across subgroups. Brier scores were higher, indicating poorer calibration, in male participants (0.20) compared to female participants (0.17), and in English-speaking participants (0.19) compared to French-speaking participants (0.17). Calibration was comparable between age groups (0.18), indicating that the predicted probabilities were equally well calibrated and that the difference reflects the position of the decision threshold relative to each group's score distribution rather than a loss of calibration.

To assess robustness to language-related distribution shifts, the model was trained on one language and evaluated on the other. The model retained discriminative performance in both directions, achieving an AUROC of 0.79 for French-to-English transfer and 0.80 for English-to-French transfer (Table~\ref{tab:cross-language}). Sensitivity remained high in both settings (0.84 and 0.88, respectively), whereas specificity was approximately 0.51 in both directions. Brier scores increased to 0.21 compared with the cross-validated performance of the full model, indicating reduced calibration under cross-language transfer. These findings suggest that the learned representations retain discriminative information across languages, but additional calibration or threshold adjustment may be required for deployment in a different linguistic population.

\begin{table}[t]
\caption{Contribution of clinical variable groups, with and without the voice modalities.}
\label{tab:clinical-groups}
\setlength{\tabcolsep}{6pt}
\centering
\begin{tabular}{lcccccccc}
\toprule
& \multicolumn{4}{c}{\textbf{Clinical}} & \multicolumn{4}{c}{\textbf{Voice + clinical}} \\
\cmidrule(lr){2-5} \cmidrule(lr){6-9}
\textbf{Clinical group} & \textbf{AUROC} & \textbf{Sens.} & \textbf{Spec.} & \textbf{Brier}
& \textbf{AUROC} & \textbf{Sens.} & \textbf{Spec.} & \textbf{Brier} \\
\midrule
Demographic & 0.56 & 0.61 & 0.45 & 0.25 & 0.71 & 0.64 & 0.63 & 0.23 \\
Lifestyle & 0.64 & 0.59 & 0.60 & 0.23 & 0.74 & 0.65 & 0.66 & 0.22 \\
Symptom & 0.71 & 0.81 & 0.34 & 0.23 & 0.73 & 0.62 & 0.68 & 0.23 \\
Treatment & 0.60 & 0.23 & 0.96 & 0.23 & 0.77 & 0.65 & 0.70 & 0.21 \\
\midrule
All clinical & 0.75 & 0.65 & 0.74 & 0.20 & 0.83 & 0.70 & 0.75 & 0.18 \\
\bottomrule
\end{tabular}
\vspace{3pt}
\parbox{1.0\linewidth}{\footnotesize Clinical variables were divided into demographic (age, sex, language, BMI), lifestyle (smoking, alcohol, perceived stress), symptom (VQ11, FSS) and treatment (antihistamine use) groups, each evaluated separately. Reported values are means across repeated 10$\times$10 stratified cross-validation folds.}
\end{table}

To further characterise the contribution of structured clinical information, clinical variables were divided into demographic, lifestyle, symptom and treatment groups, each evaluated with and without voice modalities (Table~\ref{tab:clinical-groups}). Among individual clinical groups, symptom-related variables provided the strongest discrimination when used alone (AUROC 0.71), exceeding lifestyle (0.64), treatment (0.60), and demographic variables (0.56). Adding voice modalities improved AUROC across all four clinical groups. The largest improvement was observed for treatment variables (+0.17), while the smallest was observed for symptom variables (+0.02). Voice integration also resulted in more balanced sensitivity and specificity across the clinical groups, particularly for symptom and treatment variables, which showed pronounced sensitivity-specificity trade-offs when used alone. 

\section{Discussion}\label{discussion}

This study presents a multimodal MoE framework for self-reported asthma identification that integrates acoustic embeddings from sustained vowel phonation and reading passage tasks with structured clinical and demographic features. It addresses several limitations that have constrained prior work in voice-based asthma identification. Previous studies have often relied on controlled recording conditions or specialised hardware \cite{Robin2020}, relatively small cohorts \cite{Lyu2025, Larsen2025}, or unimodal acoustic analysis without detailed clinical characterisation \cite{Kaur2023}. In contrast, the proposed framework was evaluated on a cohort of 1,218 participants using recordings acquired in real-world settings, alongside comprehensive clinical annotation including validated questionnaires assessing respiratory quality of life, fatigue, stress, smoking, and medication use — a depth of multimodal characterisation not previously reported in the asthma vocal biomarker literature.

The proposed approach achieved a mean AUROC of 0.83 across repeated 10$\times$10 stratified cross-validation folds, with adequate probabilistic calibration (Brier score 0.18) -- a level of discriminative and probabilistic performance that supports the potential utility of the proposed framework as a non-invasive screening tool. The well-calibrated probability outputs are particularly important: a screening tool that outputs miscalibrated probabilities could systematically over- or under-refer asthma cases, whereas the reliability diagram shows close agreement between predicted and observed probabilities, with only minor underestimation  of risk across the low-to-mid probability range. 

The ablation study further demonstrates that combining voice with clinical data confers systematic performance gains over any single modality. Clinical and demographic features alone achieved the strongest unimodal performance (AUROC 0.75, accuracy 0.69), consistent with the established role of BMI, smoking history, and allergic comorbidities in asthma risk stratification \cite{Zein2021, Tomisa2021, Listyoko2024}. Among the voice modalities, sustained vowel phonation outperformed reading passage speech, suggesting that phonatory stability measures derived from controlled vocal tasks may be more sensitive to respiratory dysfunction than connected speech. This observation aligns with prior work linking altered expiratory control and phonatory instability to obstructive airway disease \cite{Dogan2007}.

Bimodal combinations of voice and clinical data outperformed their constituent unimodal approaches, though with distinct performance profiles that have different clinical implications. The RP+CD combination achieved the highest bimodal AUROC (0.81) alongside the best specificity (0.79) and precision (0.77), indicating a low false positive rate that is desirable in screening contexts where unnecessary referrals carry clinical and economic costs. The SV+CD combination achieved nearly balanced sensitivity (0.70) and specificity (0.72), which may be preferable in settings prioritising case detection. The comparatively poor performance of the SV+RP combination (AUROC 0.71, accuracy 0.63), which fell below clinical features used alone, reinforce that structured clinical information provides the dominant complementary signal when fused with acoustic features.

Although the incremental AUROC improvement over the strongest bimodal model was modest and not statistically significant, the full model achieved the highest AUROC, accuracy, and F1 score, suggesting a potential benefit from integrating all three modalities even when gains in discrimination are limited. The bimodal models nevertheless offer a strong performance–efficiency trade-off, achieving comparable performance with reduced recording burden. With either voice task requiring under two minutes to collect, this supports practical deployment in remote or low-resource settings.

Comparison with conventional fusion approaches provides further context for the contribution of modality-specific processing. The MoE substantially outperformed logistic regression and multilayer perceptron, likely because modality-specific projections prevent the high-dimensional acoustic features from diluting the clinical signal. Gradient boosting was more competitive, with comparable discrimination, but the MoE achieved significantly higher sensitivity (0.70 vs. 0.65), favouring case detection. Beyond predictive performance, the MoE produces a participant-specific weight for each modality as part of the prediction process, which is not provided by concatenation-based models. This feature may be valuable in clinical settings where transparency and clinician confidence are important considerations.

These findings also have broader implications for the design of future vocal biomarker studies. Several studies have reported voice-based asthma identification in isolation \cite{Kaur2023, Lyu2025}, without integrating the clinical context that is routinely available at the point of screening. Our results suggest that such studies may underestimate the achievable performance of vocal biomarkers when appropriately contextualised with clinical information, and that the field would benefit from systematic multimodal evaluation as a standard rather than an optional extension.

The moderate fold-level instability in expert routing warrants careful interpretation. This pattern likely reflects heterogeneity in the demographic and linguistic composition of folds rather than architectural instability. Nevertheless, it highlights that the model's routing strategy is sensitive to data distribution shifts, a relevant consideration for deployment across clinical sites with different patient populations.

Correlation analyses with symptom severity provided additional exploratory insight into gating behaviour. Reading-passage gate weights increased with VQ11 score, whereas sustained-vowel weights decreased, with the same pattern observed for fatigue severity. As the VQ11 measures respiratory symptom burden \cite{Ninot2013}, this pattern may reflect the complementary information captured by the two tasks: sustained vowels primarily probe phonatory control, whereas connected speech places greater demands on respiratory coordination and continuous voice production. The shift toward reading with increasing symptom burden therefore suggests greater relative informativeness of connected speech. Because gate weights represent relative allocation of model reliance, the opposing directions of reading and sustained-vowel weights should be interpreted as a shift in task informativeness rather than opposing biological effects. The small effect sizes indicate that symptom severity explains only a limited proportion of gating variability, suggesting that the relative contribution of each task is influenced by factors beyond symptom burden.

The multilingual composition of the cohort additionally provides preliminary evidence of cross-lingual applicability. Cross-lingual transfer retained discrimination when training and testing languages were reversed (AUROC 0.79 and 0.80), but reduced specificity and increased sensitivity suggest a systematic upward shift in predicted asthma risk rather than a loss of ranking ability. This may reflect unfamiliar-language being interpreted as a deviation from training-language patterns, shifting more participants toward the asthma class. Sustained vowel phonation is on the other hand inherently language-independent, and prior work has shown that acoustic biomarkers from controlled vocal tasks generalise more robustly across languages than connected speech \cite{Despotovic24_interspeech}.

Decomposing the clinical features showed that symptom variables were the strongest individual predictors, but voice improved discrimination across all clinical groups. The gain was largest when symptom information was absent (AUROC +0.15 with demographics and +0.17 with treatment) and smallest when symptom questionnaires were included (+0.02), suggesting that voice and symptom scores capture mostly overlapping information about respiratory symptom burden. These findings support the potential of voice as a low-burden alternative to structured symptom questionnaires when these are unavailable, rather than as an addition to it. 

A consistent pattern emerged across three settings — age groups, cross-lingual transfer, and addition of voice to symptom questionnaires: discrimination remained relatively stable, while performance at the fixed decision threshold changed substantially. This indicates that model ranking is more robust than its probability scale. For potential screening applications, this distinction is important because referral depends on the decision threshold. Threshold selection should therefore be validated in the target population rather than treated as an implementation detail.

Finally, the error analysis further contextualises the model's limitations. Misclassified asthma cases tended to be older and exhibited lower VQ11 and FSS scores than correctly classified participants, suggesting that individuals with milder or atypical symptom profiles are more difficult to detect. The absence of sex-related differences in misclassification rates indicates that model errors were not systematically associated with sex. Together, these findings suggest that the proposed framework may perform less reliably in subclinical or atypical asthma presentations, which remains an important consideration for potential deployment as a general screening tool.

Several limitations of the present study should be acknowledged. Colive Voice relies on self-reported asthma diagnosis, which introduces the possibility of diagnostic misclassification. Objective confirmation through spirometry with bronchodilator reversibility testing would strengthen the validity of the ground truth labels. The study cohort was limited to adults, and the generalisability of the findings to paediatric populations, where asthma is particularly prevalent and spirometry is often impractical \cite{Hsu2010}, remains to be established. The case-control sampling and balanced class distribution do not reflect asthma prevalence in the general population and may therefore limit the direct interpretation of predictive values and performance in a clinically representative population. External validation in independent cohorts is required to assess the robustness and generalisability of the proposed model across different populations and languages, but is currently limited by the lack of multimodal asthma datasets with aligned voice modalities and rich clinical annotations. 

The present study should therefore be understood as establishing feasibility for risk prediction rather than as validating a screening tool ready for clinical deployment. Building on this foundation, future work should prioritise prospective, multi-centre validation in a clinically representative population, and the development of shared multimodal benchmarks to enable reproducible evaluation.

\section{Methods}\label{methods}

\subsection{Study design and ethics approval}\label{study}

The Colive Voice study is an international digital health initiative coordinated by the Luxembourg Institute of Health, with the goal of developing vocal biomarkers for the remote monitoring of chronic diseases and common health conditions. The study maintains a multilingual audio databank in six languages (English, French, German, Spanish, Portuguese, and Arabic) with a diverse set of standardised voice tasks: sustained vowel phonation, reading passage, spontaneous speech, picture description, and diadochokinetic task. Voice recordings are linked to rich clinical and demographic annotations collected through clinically validated, disease-specific questionnaires covering symptom burden, treatment history, and health-related quality of life. The platform is open to any individual who provides informed consent through the Colive Voice application\footnote{https://form.jotform.com/colivestudy/colive-voice}. Accordingly, all participants provided electronic informed consent prior to data collection.

The study was conducted in accordance with the principles of the Declaration of Helsinki. Ethical approval was obtained from the National Research Ethics Committee of Luxembourg (Comité National d'Éthique de Recherche, CNER) (Reference No. 202103/01; approval granted in March 2021). The study protocol was prospectively registered with \href{https://clinicaltrials.gov/study/NCT04848623?viewType=Card&term=colive%20voice&rank=1}{ClinicalTrials.gov} (Identifier: NCT04848623) on 15 April 2021. 

\subsection{Data preprocessing}\label{methods:preprocessing}

Participants were instructed to record voice samples in a quiet environment to minimise ambient noise and preserve recording quality. For each voice task, participants were provided with a reference example in their preferred language. A standardised preprocessing pipeline was applied to harmonise the format and level of recordings collected in real-world settings. Preprocessing included DC offset removal, resampling to 16 kHz, stereo-to-mono conversion, trimming of leading and trailing silence, and peak normalization.

Participant eligibility was determined based on metadata completeness, availability of the required voice tasks, presence of respiratory or neurological comorbidities known to affect voice production, and predefined audio quality-control criteria. All records were screened, with 4 excluded because of unavailable metadata. Beyond these, no clinical or demographic data was missing, and no imputation was therefore required. Among the 762 participants reporting a prior asthma diagnosis, 66 were excluded because they lacked recordings for at least one required voice task, leaving 696 participants. Task completeness did not differ significantly between participants with and without asthma (91.3\% vs. 90.1\%, $\chi^2=0.93$, $p=0.34$). A further 78 participants were excluded because of a co-reported chronic respiratory or neurological condition, and 9 were excluded due to inadequate recording quality. Recording quality was assessed using the quality control pipeline and verified by auditory review. This yielded 609 eligible asthma cases, which were subsequently matched to healthy controls on sex, age, and language.

\subsection{Feature extraction}\label{methods:features}

The two voice tasks differ fundamentally in their acoustic content and were therefore encoded using task-specific models. Sustained vowel phonation is a steady-state, non-linguistic signal in which relevant information is primarily related to phonatory quality, such as periodicity, spectral noise, and stability, with little lexical or prosodic structure. A general-purpose audio encoder such as BYOL-A \cite{Niizumi2021, Niizumi2023}, trained without linguistic supervision on diverse environmental and acoustic sounds is therefore well suited to this task, whereas speech-recognition representations are shaped by phonetic content that sustained vowels largely lack. 

Reading passages, in contrast, contain connected speech in which breath-group patterning, pauses, and articulatory dynamics may carry relevant information. These characteristics are well represented by a large-scale multilingual speech encoder, like Whisper \cite{Radford2022}. Its multilingual pretraining is also appropriate for the mixed English- and French-speaking cohort. 

\subsubsection{Sustained vowel phonation}

Acoustic embeddings from sustained vowel phonation recordings were extracted using Bootstrap Your Own Latent for Audio (BYOL-A) \cite{Niizumi2021, Niizumi2023}, a self-supervised pretraining framework. The model takes as input 96$\times$64 log-Mel spectrograms computed from preprocessed audio recordings. It further creates two augmented versions of each spectrogram by shifting pitch and stretching time, and outputs 2048-dimensional embeddings, which are further passed through online and target networks. The online network aims to predict the representation of the target network, whose parameters are updated progressively using the exponential moving average of the online network’s weights. 

\subsubsection{Reading passage}

Acoustic embeddings for reading passage recordings were extracted using the encoder of the Whisper Large V3 model, a transformer-based automatic speech recognition system pretrained on large-scale multilingual and multitask speech data \cite{Radford2022}. Preprocessed audio recordings were converted into log-Mel spectrograms using the Whisper feature extraction pipeline. The encoder was applied in inference mode with frozen weights to obtain contextualized frame-level representations with dimensionality 1280. Utterance-level embeddings were generated by mean pooling across the temporal dimension, resulting in a single fixed-length representation per recording.

\subsubsection{Clinical and demographic features}

Structured clinical and demographic features were extracted for each participant from the Colive Voice dataset. The following variables were included: age, sex, language, BMI, smoking status, alcohol consumption, antihistamine use, perceived stress level, respiratory quality of life score (VQ11) \cite{Ninot2013}, and fatigue severity scale score (FSS) \cite{Krupp1989}, as shown in Table~\ref{tab:cohort}. Continuous variables (age, BMI, VQ11, and FSS scores) were used as measured. BMI was modelled as a continuous value, with the WHO categories in Table~\ref{tab:cohort} shown for descriptive purposes only. Binary variables (sex, language, and antihistamine use) were represented as 0/1 indicators. For binary variables, this is equivalent to one-hot encoding. Ordinal variables (smoking status, alcohol consumption, and perceived stress) were integer-encoded according to the response ordering of the original questionnaire items, thereby preserving their ordinal structure. These variables were therefore not treated as nominal categories, but as ordered predictors.

\subsection{Model architecture}
\label{methods:architecture}

We developed a Mixture-of-Experts framework for multimodal self-reported asthma identification that integrates acoustic embeddings from two standardised voice tasks — sustained vowel phonation and reading passage — with structured clinical and demographic features. The architecture consists of three modality-specific components: (i) feature projection layers, (ii) modality-specific expert networks, and (iii) hierarchical gating mechanisms operating at both the expert and modality-group levels.

Each modality input is first transformed into a lower-dimensional latent representation using a dedicated linear projection layer followed by layer normalisation. Sustained vowel and reading passage embeddings were projected to 64 dimensions, and the clinical and demographic vector to 32 dimensions. No non-linear activation is applied after projection in order to preserve the structure of the pretrained embedding space while reducing dimensionality and improving optimisation stability.

The projected representations are processed by independent expert networks specialised for each modality. Four experts are assigned to sustained vowel phonation (VE1--VE4), four to reading passage (RE1--RE4), and two to clinical and demographic features (CE1--CE2), yielding ten experts in total. Each expert consists of a single hidden layer with 128 units and ReLU activation followed by a linear output layer producing a scalar logit.

To adaptively combine expert outputs, the model employs a hierarchical gating strategy. First, an expert-level gating network determines the contribution of individual experts within and across modality groups. The gate receives as input the concatenation of modality-specific gate embeddings with dimension $3 \times 128 = 384$, passed through a two-layer gating network that produces logits for all experts. Gate probabilities are computed using a temperature-scaled softmax:

\begin{equation}
g_i = \frac{\exp(z_i / T)}{\sum_j \exp(z_j / T)}
\end{equation}
where $z_i$ denotes the logit associated with expert $i$ and $T$ is a temperature parameter controlling gate sharpness. The temperature is annealed during training from higher to lower values to promote exploration of experts early in training and progressively encourage more specialized routing as learning stabilises.

\begin{equation}
T_{\text{epoch}} = \max(0.5,\ 1.0 \cdot 0.98^{\text{epoch}})
\end{equation}
starting near $T = 1.0$ and decaying towards a lower bound of $T = 0.5$, progressively sharpening the gate distribution over training. Lower values of $T$ produce sharper, more confident expert assignments, whereas higher values yield softer distributions that encourage more uniform expert participation during training. During evaluation, temperature was fixed at $T = 1.0$ to reflect the model's learned gate distribution without artificial sharpening.

\begin{table}[t]
\caption{Trainable parameter counts for each model configuration.}
\label{tab:params}
\setlength{\tabcolsep}{15pt}
\centering
\begin{tabular}{lccc}
\toprule
\textbf{Configuration} & \textbf{Architecture} & \textbf{Experts} & \textbf{Trainable parameters} \\
\midrule
SV & MLP & --- & 557,569 \\
RP & MLP & --- & 360,961 \\
CD & MLP & --- & 35,841 \\
\midrule
SV+RP & MoE & 8 & 332,050 \\
SV+CD & MoE & 6 & 220,910 \\
RP+CD & MoE & 6 & 171,758 \\
\midrule
SV+RP+CD & MoE & 10 & 362,679 \\
\bottomrule
\end{tabular}
\vspace{3pt}
\parbox{1.0\linewidth}{\footnotesize Unimodal models use a Multilayer Perceptron with a 256--128--1 architecture, whereas multimodal models use the Mixture-of-Experts architecture. Both audio encoders were frozen, so the counts refer only to the layers that were trained.}
\end{table}

Second, a modality-level gating mechanism adaptively weights the aggregated contribution of each modality group. This hierarchical formulation was introduced to explicitly separate expert selection from modality importance estimation. In multimodal clinical data, modalities may differ substantially in predictive relevance across individuals due to heterogeneity in symptom manifestation, recording quality, or missing clinical signal. The group-level gate therefore enables the model to dynamically adjust the relative contribution of sustained vowel phonation, reading passage, and clinical features at the individual level, while preserving specialisation among experts within each modality.

Within each modality group, expert outputs are combined using their corresponding expert-level gate probabilities. The final prediction is then computed as a weighted combination of modality-group mixtures:

\begin{equation}
\hat{y} =
\sum_{j \in {\mathrm{SV,RP,CD}}}
w_j
\left(
\sum_{i \in \mathcal{E}_j}
g_i \hat{y}_i
\right)
\end{equation}
where $w_j$ denotes the modality-level gate weight for modality group $j$, $g_i$ denotes the expert-level gate probability for expert $i$, and $\hat{y}_i$ represents the scalar output of expert $i$, and $\mathcal{E}_m$ denotes the subset of experts assigned to modality group $j$ within the hierarchical MoE architecture.

In total, the full multimodal model has 362,679 trainable parameters, comprising the projection layers, the ten experts and the two gating networks. Parameter counts for all configurations are given in Table~\ref{tab:params}. Notably, the unimodal models are larger than the multimodal ones despite processing fewer inputs. This follows from the corresponding architectures: MLP passes the full 2,048- or 1,280-dimensional embeddings for sustained vowel phonation and reading passage, respectively, directly into a 256-unit hidden layer, whereas the Mixture-of-Experts first projects each modality to 64 dimensions, so its experts and gating networks operate in a much smaller space. Differences in performance across configurations are consequently not attributable to model capacity.

\subsection{Training procedure}
\label{methods:training}

Models were trained using the Adam optimiser with a learning rate of $10^{-4}$ and a batch size of 64 for 30 epochs. Binary cross-entropy with logits was used as the loss function. We used Python (version 3.11) and PyTorch (version 2.12, CUDA 12.8) for experiments in this study.

Both audio encoders were used as fixed feature extractors and were kept frozen throughout model training. BYOL-A and the Whisper encoder were run in inference mode without gradient updates, and the resulting embeddings were extracted once and cached prior to downstream model training. Given the cohort size of 1,218 participants, fine-tuning encoders of this scale was considered likely to increase the risk of overfitting without providing sufficiently reliable improvements in representation quality.



Each model was trained for a fixed epoch budget and evaluated once, at the final epoch, on the held-out fold, ensuring that the test fold did not influence model selection. 
The final-epoch model was used to generate fold-level predictions and gate weights.

\subsection{Evaluation}
\label{methods:evaluation}

Model performance was evaluated using repeated 10-fold stratified cross-validation with 10 repetitions ($10 \times 10$ CV), following the recommendations of de Hond et al.~\cite{deHond2023} for clinical prediction model validation. Stratification preserved class balance across folds. Splits were fixed prior to training and held constant across all model configurations to ensure fair comparison in the ablation study.

Performance metrics, including AUROC, accuracy, sensitivity, specificity, precision, F1 score, and Brier score, were computed across all folds. Threshold-dependent metrics were computed at a fixed probability threshold of 0.5, specified a priori and held constant across all model configurations and subgroup analyses. No threshold optimisation was performed on held-out data.

To account for the non-independence of folds within the same repetition, summary statistics were computed by first averaging metrics within each repetition, and further computing the mean and 95\% confidence intervals across the 10 repetition-level means. Calibration was assessed using reliability diagrams with 8 uniform bins and the Brier score.

\subsection{Modality-level routing analysis}
\label{methods:interpretability}

Expert-level and group-level gate weights were recorded for each validation participant during evaluation. Since each participant appears in the validation set once per repetition, both expert-level gate weights (dimension: 
$n_{\text{experts}}=10$) and modality group-level gate weights (dimension: $n_{\text{group}}=3$, 
corresponding to sustained vowel phonation, reading passage, and clinical/demographic modalities) were accumulated across all 10 repetitions and averaged per participant independently, yielding stable per-individual estimates for both levels of the gating hierarchy. The statistical unit in all analyses is therefore the participant. Expert-level gate weights were averaged across fold positions within each cross-validation fold for heatmap visualisation. Class-stratified gate weight distributions were compared between self-reported asthma cases and healthy controls using two-sided Mann-Whitney U tests ($n=609$ per group, one averaged value per participant). Spearman rank correlations were computed between modality-group gate weights and clinical scores (VQ11 and FSS) in self-reported asthma cases only ($n=609$).

Error analysis was performed by assigning true positive and false negative cases using majority vote across all 10 repetitions of cross-validation. This was used to assess whether clinical characteristics differ between correctly classified and misclassified asthma cases. Clinical profiles of the two groups were compared using two-sided Mann-Whitney U tests for continuous variables (VQ11, FSS, age) and chi-square test for sex.

\subsection{Ablation study and baseline models}\label{methods:ablation}

To quantify the contribution of each modality, unimodal and bimodal model configurations were trained and evaluated under identical experimental conditions, using the same hyperparameters, training procedure, and cross-validation splits as the full model. Performance was compared across seven configurations (SV, RP, CD, SV+RP, SV+CD, RP+CD, and SV+RP+CD).

To assess whether the MoE architecture provides an advantage over conventional multimodal fusion, three baseline models were fitted to the concatenation of all three modality feature vectors: L2-regularised logistic regression, eXtreme Gradient Boosting (XGBoost) \cite{Chen2016}, and MLP. All baselines used the same cross-validation folds as the MoE, enabling paired fold-wise comparisons. For logistic regression and MLP, standardisation was performed within a pipeline fitted separately on the training data of each fold and each inner split during hyperparameter selection, preventing information leakage. XGBoost is scale invariant and was therefore fitted without standardisation.

The MLP used the same 256--128--1 architecture, learning rate ($10^{-4}$), batch size of 64, and fixed 30-epoch budget as the MoE. Logistic regression regularisation was selected within each training fold using three-fold inner cross-validation over $C \in \{0.001, 0.01, 0.1, 1\}$, optimising AUROC. XGBoost used the library default hyperparameters with early stopping after 15 rounds without improvement on a 10\% stratified validation split of the training fold. The held-out test fold was never used for model selection. The models were implemented in scikit-learn (version 1.8) and xgboost libraries (version 3.2).

\subsection{Statistical analysis}\label{methods:stats}

All statistical analyses were performed in Python using SciPy (version 1.17). Confidence intervals were computed using the $t$–distribution. Group comparisons for continuous variables were performed using two-sided Mann–Whitney U test, while categorical variables were compared using chi-square test. Spearman rank correlation was used for monotonic association between continuous variables. Where multiple comparisons were performed, $p$–values were corrected using the Benjamini–Hochberg false discovery rate method. A significance threshold of $p<0.05$ was applied throughout.

\backmatter

\section*{Funding}
The Colive Voice study is supported by internal funding from the Luxembourg Institute of Health. The funder had no role in the study design, data collection, data analysis and interpretation, decision to publish, or preparation of this manuscript.

\bibliography{sn-bibliography}

\end{document}